\documentclass[fleqn,usenatbib]{mnras}

\usepackage{newtxtext,newtxmath}
\usepackage{mathptmx}
\usepackage{txfonts}

\usepackage[T1]{fontenc}
\usepackage{ae,aecompl}

\usepackage[figuresright]{rotating}
\usepackage{epsfig}
\usepackage{graphics}
\usepackage{epstopdf}
\usepackage[latin1]{inputenc}
\usepackage{latexsym}

\usepackage{graphicx}	
\usepackage{amsmath}	
\usepackage{amssymb}	

\title[Type Iax Supernova 2014ek]{Optical Observations of the 2002cx-like Supernova 2014ek, and Characterizations of SNe Iax}

\author[Li et al.]{Linyi Li$^{1}$, Xiaofeng Wang$^{1}$, Jujia Zhang$^{2,3,4}$, Iair Arcavi$^{5,6}$, Tianmeng Zhang$^{7,8}$,
\newauthor Liming Rui$^{1}$, Griffin Hosseinzadeh$^{5,6}$, D. Andrew Howell$^{5,6}$, Curtis McCully$^{5,6}$,
\newauthor Kaicheng Zhang$^{1}$, Stefano Valenti$^{9}$, Jun Mo$^{1}$, Wenxiong Li$^{1}$, Fang Huang$^{10,1}$,
\newauthor Danfeng Xiang$^{1}$, Lifan Wang$^{11,12}$, and Xu Zhou$^{8}$
\\
$^{1}$Physics Department and Tsinghua Center for Astrophysics (THCA), Tsinghua University, Beijing, 100084, China; wang\_xf@mail.tsinghua.edu.cn\\
$^{2}$Yunnan Observatories (YNAO), Chinese Academy of Sciences, Kunming 650216, China\\
$^{3}$Key Laboratory for the Structure and Evolution of Celestial Objects, Chinese Academy of Sciences, Kunming 650216, China\\
$^{4}$Center for Astronomical Mega-Science, Chinese Academy of Sciences, 20A Datun Road, Chaoyang District, Beijing, 100012, China\\
$^{5}$Las Cumbres Observatory, 6740 Cortona Dr Suite 102, Goleta, CA 93117-5575, USA\\
$^{6}$Department of Physics, University of California, Santa Barbara, CA 93106-9530, USA\\
$^{7}$Key Laboratory of Optical Astronomy, National Astronomical Observatories of China, Chinese Academy of Sciences, Beijing 100012, China\\
$^{8}$School of Astronomy and Space Science, University of Chinese Academy of Sciences, Beijing 101408, China\\
$^{9}$Department of Physics, University of California, Davis, CA 95616, USA\\
$^{10}$Department of Astronomy, Shanghai Jiao Tong University, Shanghai, 200240, China \\
$^{11}$ George P. and Cynthia Woods Mitchell Institute for Fundamental Physics $\&$ Astronomy, Texas A. $\&$ M. University, College Station, TX 77843, USA\\
$^{12}$ Purple Mountain Observatory, Chinese Academy of Sciences, Nanjing, 210034, China}

\pubyear{2017}

\begin{document}
\label{firstpage}
\pagerange{\pageref{firstpage}--\pageref{lastpage}}
\maketitle

\begin{abstract}

We present optical observations of supernova (SN) 2014ek discovered during the Tsinghua-NAOC Transient Survey (TNTS), which shows properties that are consistent with those of SN 2002cx-like events (dubbed as SNe Iax). The photometry indicates that it is underluminous compared to normal SNe Ia, with the absolute $V$-band peak magnitude being as $-17.66\pm0.20$ mag. The spectra are characterized by highly ionized Fe {\sc iii} and intermediate-mass elements (IMEs). The expansion velocity of the ejecta is found to be $\sim$5000 km s$^{-1}$ near the maximum light, only half of that measured for normal SNe Ia. The overall spectral evolution is quite similar to SN 2002cx and SN 2005hk, while the absorption features of the main IMEs seem to be relatively weaker. The ${}^{56}$Ni mass synthesized in the explosion is estimated to be about 0.08 M$_{\odot}$ from the pseudo bolometric light curve. Based on a large sample of SNe Iax, we examined the relations between peak luminosity, ejecta velocity, decline rate, and peak $V - R$ color but did not find noticeable correlations between these observables, in particular when a few extreme events like SN 2008ha are excluded in the analysis. For this sample, we also studied the birthplace environments and confirm that they still hold the trend of occurring preferentially in late-type spiral galaxies. Moreover, SNe Iax tend to occur in large star-forming regions of their host galaxies, more similar to SNe Ibc than SNe II, favoring that their progenitors should be associated with very young stellar populations.

\end{abstract}


\section{Introduction}

Supernovae (SNe) represent the final, explosive stage in the evolution of certain types of stars, playing important roles in diverse aspects of astrophysics. With the dramatic increase in the number of SNe discovered over the past decade, numerous types and subtypes are introduced to describe their diverse observational properties. In theory, SNe have been considered to form in two main channels: thermonuclear explosions of accreting white dwarf (WD) in a binary system and core-collapse (CC) explosions of massive stars. How different types of SNe are produced from different types of stars is not yet well understood. An accurate physical model of SNe is important not only for our understanding of late stages of stellar evolution but also for precision cosmology.

Among the thermonuclear WD explosions, normal type Ia supernovae (SNe Ia) belong to the most common subclass and are well known as distance indicators to measure the expansion rate of the Universe (Riess et al. 1998; Perlmutter et al. 1999). They are generally believed to arise from explosions of accreting carbon-oxygen (CO) WD with a mass close to the 1.4 M$_{\odot}$ in a binary system, although large scatter emerge in their observed properties. For example, some SNe Ia show weak features of intermediate-mass elements (IMEs) and prominent Fe {\sc ii}/Fe {\sc iii} lines like the luminous subclass of SN 1991T (Filippenko et al. 1992), while others show strong features of IMEs and prominent Ti {\sc ii} lines like the faint subclass of SN 1991bg (see reviews by Filippenko 1997). Aside from these subtypes, some exotic events of thermonuclear explosion also emerge in recent years as a result of the expansion of wide-field survey projects. One weird subclass has very high luminosity and strong absorption of unburned carbon in their spectra (Howell et al. 2006). Representative example include SN 2007if (Scalzo et al. 2010) and SN 2009dc (Yamanaka et al. 2009), which could be due to explosions of super-chandrasekhar-mass WD (i.e., with a mass $\gtrsim$ 2.0 M$_{\odot}$). The SN 2002cx-like SN represents another weird subclass, which was initially noticed by Li et al. (2003) and is characterized by an SN 1991T-like pre-maximum spectrum (i.e., highly ionized Fe {\sc iii}), an SN 1991bg-like luminosity and low ejecta velocities.

There are currently about 45 SNe that have been identified as SN 2002cx-like subclass (also dubbed as SNe Iax) according to their spectral features. Compared to normal SNe Ia, SNe Iax are found to show obviously weaker absorptions of IMEs in their spectra and relatively stronger spectral features of iron-group elements (IGEs) in the early phase. The expansion velocities measured from the spectra of SNe Iax are found to be only roughly half those of normal SNe Ia, ranging from $\sim$2000 km s$^{-1}$ to $\sim$8000 km s$^{-1}$.
At late times, the spectra of normal SNe Ia are dominated by broad forbidden emission lines of IGEs, while SNe Iax show narrow permitted lines of IGEs (Jha et al. 2006). Photometrically, SNe Iax are found to be remarkably fainter than normal SNe Ia, with an absolute peak magnitude ranging from $\sim-13$ mag to $\sim-18$ mag in the $V$ band (Foley et al. 2013). Nevertheless, these low-luminosity SNe Iax do not show a faster decline rate (i.e., with a typical value of $\Delta$m$_{15}$(B)$\approx$1.3 mag), as expected from the Lira-Phillips relation found for normal SNe Ia (Phillips 1993). This indicates that SNe Iax have different energy sources, explosion mechanisms and/or progenitor systems compared to normal SNe Ia. Besides, two SNe Iax (SN 2004cs and SN 2007J) are found to show He {\sc i} emission line in their spectra (Foley et al. 2013), suggesting that they might have more massive progenitor systems than SNe Ia. Nevertheless, the nature
of the progenitor systems of SNe Iax is still highly debated.

Theoretically, the peculiar subclass of SNe Iax may arise from less violent explosion of a C/O WD that have accreted He-rich material
from a companion star, and this could correspond to a pure deflagration when the SN fails to change from deflagration to detonation in
the explosion (Phillips et al. 2007; Foley et al. 2009; Jordan et al. 2012; Kromer et al. 2013; and Liu et al. 2015). As an alternative,
SNe Iax have also been proposed to result from core collapse explosion of massive stars but with significant fallback onto a
new-formed blackhole (i.e., Valenti et al. 2009; Moriya et al. 2010), which is favored by the fact that SNe Iax prefer to occur in
late-time spiral galaxies and are associated with young stellar environments (Foley et~al. 2009; Lyman et al. 2013). Analysis of the
pre-explosion HST image, which may be the most effective way to constrain the progenitors of SNe, indicates that SN 2012Z may have
a progenitor system consisting of a white dwarf and a companion of non-degenerate helium-star (Yamanaka et al. 2015; McCully et al. 2014). Note, however, that current members of SNe Iax show large differences in luminosity and spectral features, suggesting that these
peculiar subclass may also come from a diverse set of progenitor systems or explosion models. Some members of current sample of
SNe Iax could be misclassified. A larger sample of SNe Iax is thus needed to address the observed diversities and their origins.

In this work, we present the observation and study of another member of 2002cx-like supernova, SN 2014ek. Some statistical results based on a collection of current SNe Iax sample are also presented. This paper is organized as follows. We describe our observations and data reductions in \S 2. Analysis of reddening, photometry, and spectra of SN 2014ek are presented in \S 3. In \S 4, we discussed the
luminosity, its correlation with photometric and spectroscopic parameters, and the progenitor environments of SN 2014ek along with
other sample. Finally, we give a brief summary in \S 5.

\section{Observation and Data Reduction}

\begin{figure}
	\includegraphics[width=\columnwidth]{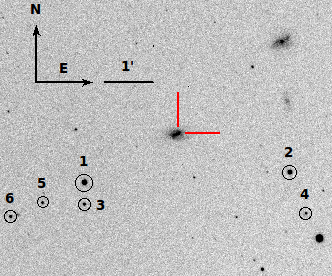}
    \caption{SN 2014ek in UGC 12850. This is a $R$-band image taken by the Lijiang 2.4 telescope on Oct. 31.63 UT 2014. The supernova is located at the center of cross-hairs and six standard stars are marked. North is up and east is to the right.}
    \label{fig:fig-1}
\end{figure}

SN 2014ek was discovered on 2014 October 18.81 UT during the Tsinghua-NAOC Transient Survey (TNTS, see Zhang et al. 2015 and Yao et al. 2015). It exploded in the spiral galaxy UGC 12850, which seems to be a Sb galaxy from its SDSS images, with a relatively less prominent barred structure, as shown in Fig.~\ref{fig:fig-1}. The coordinate of this SN is $\alpha$ = 23$^h$56$^m$06.55$^s$ and $\delta$ = $+29\degr 22 \arcmin 42.3 \farcs$, approximately 7.0 $\arcsec\,$East and 1.0 $\arcsec\,$ North from the center of the host galaxy. The redshift of UGC 12850 is 0.023, consistent with that derived from the narrow H$_{\alpha}$ emission seen in the SN spectra, which corresponds to a distance modulus of 34.99$\pm$0.20 mag (Tully et al. 2013). A spectrum taken on 2014 October 22.76 UT with the 2.4-m telescope at LiJiang Gaomeigu Station of Yunnan Observatories matches with the peculiar supernova SN 2002cx (Li et al. 2003) at about one week before the maximum light. An expansion velocity of about 6000 km s$^{-1}$ is inferred from the Si {\sc ii} 6355 absorption minima (Zhang \& Wang 2014). As SN 2014ek was discovered at a relatively early phase, we triggered the followup optical observations of this object, spanning the phase from about 6 days before to about 1 month from the $B$-band maximum light.

\subsection{Optical Photometry}

The optical photometry of SN 2014ek was collected using the 0.8-m Tsinghua University-NAOC telescope (hereafter TNT) at Xinglong Observatory of NAOC, the Las Cumbres Observatory Global Telescope 1-m network (hereafter LCO), and the Lijiang 2.4-m telescope (LJT) of Yunnan Astronomical Observatories. The TNT and LJT observations were obtained in standard Johnson-Cousin $UBVRI$ bands, while the LCO photometry was taken in the Johnson $BV$ and Sloan $gri$ filters. These observations covered the phases from October 22 2014 to November 29 2014.

All CCD images were pre-processed using standard \textsc{IRAF}\footnote{IRAF is distributed by the National Optical Astronomy Observatories, which are operated by the Association of Universities for Research in Astronomy, Inc., under cooperative agreement with the National Science Foundation (NSF).} routines, including corrections for bias, flat field, and removal of cosmic rays. As SN 2014ek locates close to the center of UGC 12850, measurements of the SN flux will be affected by the galaxy light. Thus template subtraction technique is applied to the observed images of SN 2014ek before performing photometry. The galaxy templates for the LCO, LJT, and TNT data were taken on 2016 Jun. 07.40, 2015 Sept. 12.63, and 2015 Nov. 13.15, respectively, which corresponds to 588.01 days,
319.24 days, and 381.27 days after the $B$-band maximum light. The instrumental magnitudes of both the SN and the reference stars were
then measured from the subtracted images using the aperture photometry.

To convert the instrumental magnitudes to the standard system, 6 reference stars marked in Fig.~\ref{fig:fig-1} were used to determine the photometric zeropoints. The color terms of the LCO, TNT, and LJT and the extinction coefficients at the corresponding sites are taken from Valenti et al. (2016), Huang et al. (2012), and Zhang et al. (2014), respectively. The $UBVRIugri$-band magnitudes of the field stars are listed in Table~\ref{tab:table 1}. The Johnson $UBVRI$ magnitudes of these comparison stars are transformed from the $ugri$ magnitudes from the Sloan Digital Sky Survey (SDSS) Data release 9 catalog (Ahn et al. 2012). The LCO gri-band magnitudes are also converted to the $BVRI$-band values to increase the data sampling in these wavebands. The transformation of the $UBVRI$ magnitudes from the SDSS $ugri$ magnitudes is based on the empirical formula given in Jordi et al. (2006). The final calibrated magnitudes of SN 2014ek are presented in Table~\ref{tab:table 2}. We did not apply additional magnitude corrections to the photometry from different telescopes (i.e., $S$-corrections; Stritzinger et~al. 2002) as the $BVRI$-band magnitudes obtained from the three photometric systems are overall consistent with each other (i.e., within $\sim$0.05 mag) and accurate S-corrections between different systems need detailed information about the transmission curves of different filters which are not available for the TNT system.

\begin{figure}
	\includegraphics[width=\columnwidth]{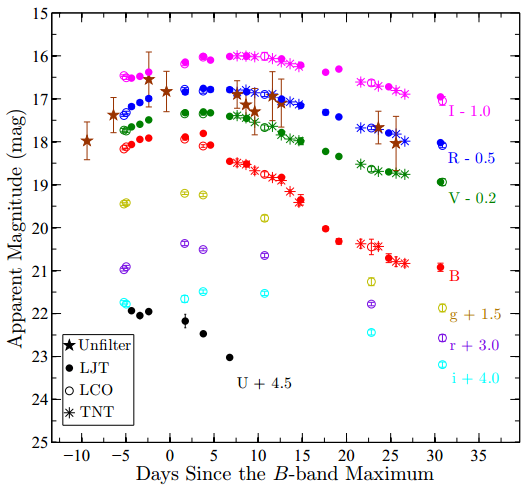}
    \caption{The $UBVRI$- and $gri$-band light curves of SN 2014ek from three different telescopes. Data from LJT 2.4-m, LCO 1-m, and TNT 0.8-m telescopes are plotted with solid points, open circles, and asterisks, respectively. The unfiltered magnitudes from the TNTS are also overplotted, which are also subtracted by 0.5 mag as the $R$-band data.}
    \label{fig:fig-2}
\end{figure}

\subsection{Optical Spectroscopy}

A total of nine low-resolution optical spectra were obtained for SN 2014ek, covering the phases from t$\sim$$-$5.1 days to +25.3 days relative to the $B$-band maximum light. Seven of these spectra were obtained with the YFOSC system mounted on Lijiang 2.4-m telescope, and fluxes of these spectra were corrected with photometric results from the LJT. Another two spectra were obtained with the FLOYDS mounted on the 2.0-m Faulkes Telescope North telescope (FTN). A journal of the spectroscopic observations and our instrument configuration are given in Table~\ref{tab:table 3}.

All spectra were reduced using the standard IRAF routines, which involves corrections for bias, flat field, and cosmic rays. The wavelength calibration is obtained by deriving the dispersion solution using FeAr and FeNe arc lamp spectra. Flux calibration was derived using the instrumental sensitivity curves of spectro-photometric standard stars observed at similar airmass on the same night as the supernova, and corresponding standard stars are also listed in Table~\ref{tab:table 3}. The spectra were further corrected for continuum atmospheric extinction during flux calibration, using mean extinction curves obtained at Lijiang Observatory in Yunnan and Haleakala Observatory in Hawaii; moreover, telluric lines were removed from the data. The prominent narrow H {\sc i} emission lines of its host galaxy has also been removed after identifying it by using low-order polynomial to fit the continuum at nearly 6500$\sim$6600\AA\ at the rest-frame wavelength.

\section{Analysis}

\subsection{Reddening}

There are two main methods to determine the host-galaxy reddening of SNe Ia: one is the photometric method, based on the correlation found between light-curve shape and the intrinsic color (Phillips et al. 1999; Wang et al. 2009; Burns et al. 2014); the other is through the correlation between the equivalent width (EW) of Na {\sc i} D absorption feature and the reddening (Munari \& Zwitter 1997; Turatto et al. 2003; Poznanski et al. 2012). Considering that SNe Iax show large scatter in their observed properties, the photometric method derived from SNe Ia cannot be applied to estimate the host-galaxy reddening for SN 2014ek. On the other hand, the spectra of SN 2014ek do not show obvious Na {\sc i} D absorption, we thus ignore the host-galaxy extinction and only take into account the Galactic component in the calculation of luminosity. For SN 2014ek, the Galactic reddening is found to be E($B-V$)=0.054 (Schlafly \& Finkbeiner 2011), corresponding to a $V$-band extinction of 0.17 mag with the Cardelli et al. (1989) extinction law (i.e., R$_{V}$=3.1).

\begin{figure*}
	\includegraphics[width=17cm]{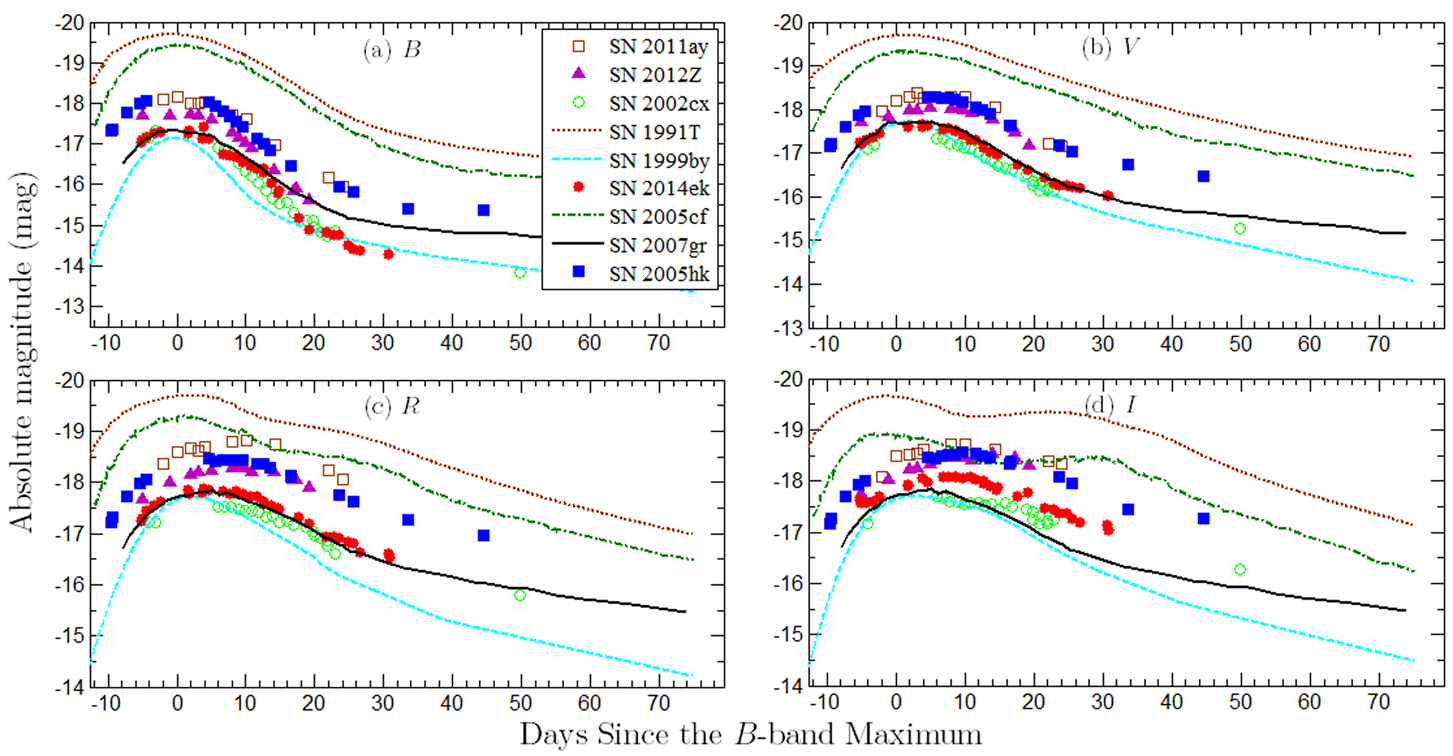}
    \caption{The $BVRI$-band light curves of SN 2014ek, compared with those of other Iax supernovae such as SNe 2002cx (Li et al. 2003), 2005hk (Phillips et al. 2007), 2011ay (Szalai et al. 2015), and 2012Z (Yamanaka et al. 2015). Different subclasses of SN Ia
    such as SN 1991T (Lira et al. 1998), SN 1999by (Bonanos et al. 1999), SN 2005cf (Wang et al. 2009) are overplotted. The light curves
    of type Ic supernova 2007gr (Chen et al. 2014) are also plotted for comparison. Note that these light curves have been corrected for the extinctions of the Milkyway and the host galaxies whenever possible.}
    \label{fig:fig-3}
\end{figure*}

\subsection{Light Curves}

Fig.~\ref{fig:fig-2} shows the $UBVRI$- and $gri$-band light curves of SN 2014ek. These light curves cover the phase from about t$\sim$-5.2 days to t$\sim$+31.0 days from the $B$-band maximum light ($B_{max}$). In comparison, the unfiltered observation started four days earlier than the multicolor photometry.

Applying low-order polynomial fits to the light curves near the maximum light yields $m_{\rm B}$(max)=17.89 $\pm$ 0.12 mag on MJD
56,958.39 $\pm$ 0.34 (2014 Oct. 28.39) and $m_{\rm V}$(max)=17.50 $\pm$ 0.04 mag on MJD 56,960.89 $\pm$ 0.18 (2014 Oct. 30.39)
This corresponds to absolute peak magnitudes as M$_{\rm B}$(max) =$-$17.32$\pm$0.23 mag and M$_{\rm V}$(max)=$-$17.66$\pm$0.20 mag
if a distance modulus of $34.99\pm0.20$ mag and an Galactic extinction of A$_{V}$= 0.17 mag are adopted in the calculation.
This indicates that SN 2014ek is apparently fainter than normal SNe Ia but comparable to SN 2002cx and SN 2005hk. Fitting
results of the peak apparent magnitude, decline rate within 15 days after the peak, and maximum-light date $t_{max}$ in
different bands are reported in Table~\ref{tab:table 4}. The absolute peak magnitudes, corrected for the Galactic extinctions,
are also presented.

In Fig.~\ref{fig:fig-3}, we compare the light curves of SN 2014ek with other well observed SNe Iax, including SN 2002cx (Li et al. 2003), SN 2005hk (Phillips et al. 2007), SN 2011ay (Szalai et al. 2015), and SN 2012Z (Yamanaka et al. 2015). The $BVRI$-band light curves of three representative subclasses of SNe Ia, such as SN 1991T (Lira et al. 1998), SN 1999by (Bonanos et al. 1999), and SN 2005cf (Wang et al. 2009), are overplotted for comparison. The light curves of SN 2007gr (Chen et al. 2014), a typical type Ic supernova, are also overplotted. One can see that the overall light curve evolution of SN 2014ek is similar to that of other comparison SNe Iax (especially SN 2002cx), and it also shows close resemblances to SN 2007gr in the $VRI$ bands.

We notice that the $B$-band light curve of SN 2014ek exhibits an abnormal post-maximum evolution, with an apparent break occurring at about 10 days from the maximum light. After t$\sim$+10 days, the $B$-band magnitude shows a faster decline for about one week relative to other comparison SNe including the subclass of SNe Iax. Such a photometric evolution is also consistent with the rapid flux
drop of the continuum on the blue end during this phase (see Section 3.4). This could be explained with a rapid decrease of opacity with
the expanding of the ejecta. As such a sudden change in the $B$-band magnitude decline was not ever seen in other SNe Iax, the fast flux
drop might be also due to the newly formed dust in the ejecta.

Unlike normal SNe Ia, SN 2014ek doesn't show a prominent secondary maximum feature in the $R$ or the $I$ band, as similarly seen
in other SNe Iax. The formation of secondary peak in the $RI$-band light curve of normal SNe Ia may be related to the opacity or
recombination effect (Pinto \& Eastman 2000; Kasen 2006). At t$\sim$ 20-30 days from the peak, the electrons recombine with the ions
like Fe {\sc iii} and Co {\sc ii} in the ejecta as the photosphere recedes, producing additional emission at this phase and this leads to
the formation of secondary peak in the longer wavebands. For SNe Iax, however, the spectra evolve fast into the nebular phase because of
rapidly decreasing photospheric temperature $T_{eff}$ and photospheric radius $R_{ph}$ soon after the maximum light. As a result, the
recombination from Fe {\sc iii} to Fe {\sc ii} occurs in relatively early phase. And this may be the reason that SNe Iax show somehow
broader peaks in the $RI$ bands. SN 2002cx seems to be an outlier in the I-band evolution if compared with other SNe Iax, showing a
plateau phase of about 20 days (Li et al. 2003), which may result from the mixing of the primary and secondary peaks
at a certain level.

\begin{figure}
	\includegraphics[width=\columnwidth]{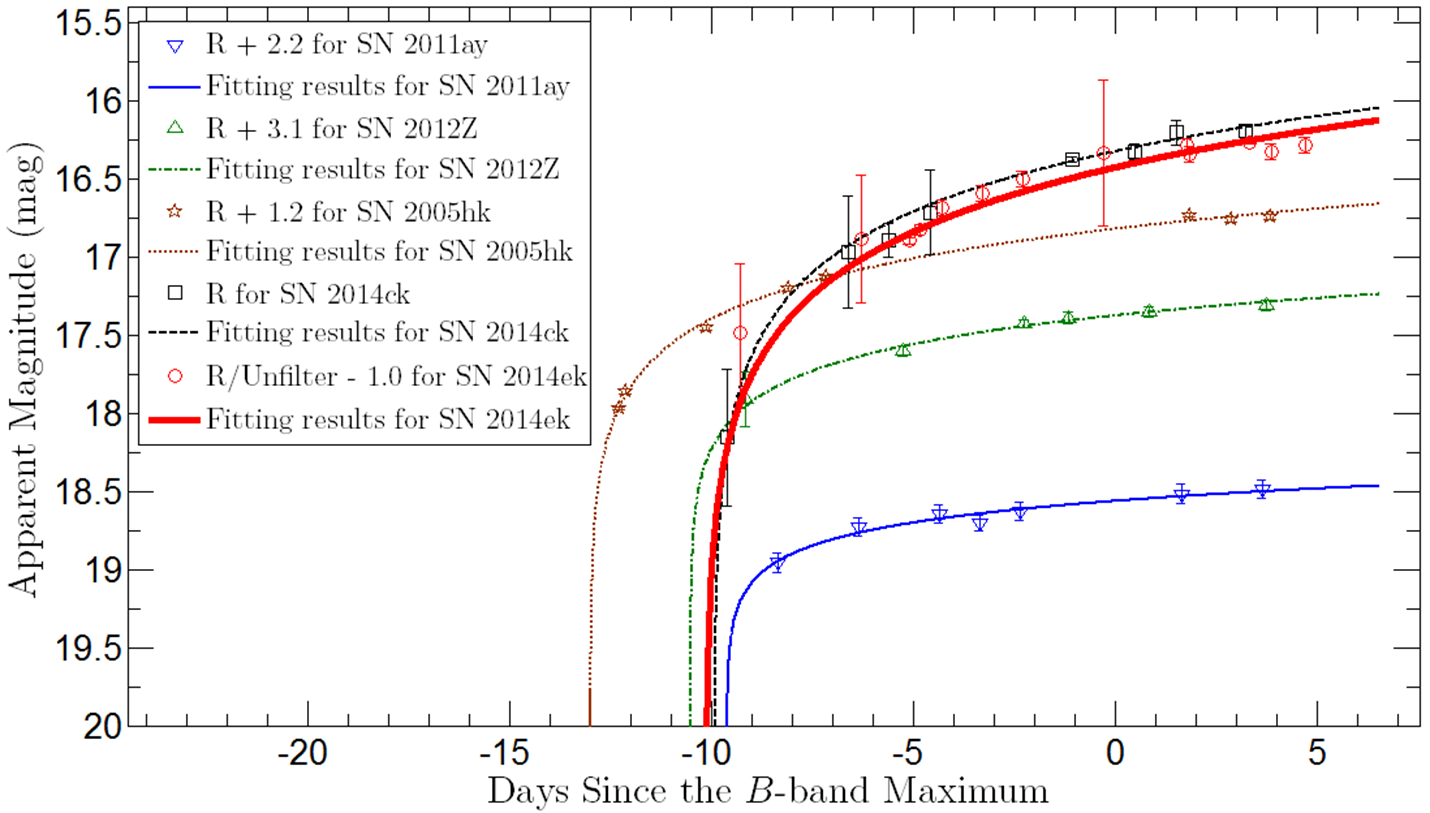}
    \caption{The $f\propto(t-t_0)^n$ model fit to the rising phase of the unfiltered and $R$-band data (open circles).
    Best-fit curve is plotted in bold solid line for SN 2014ek. Fitting to the rising evolution of some SNe Iax, including SN 2005hk
    (Phillips et al. 2007), SN 2011ay (Szalai et al. 2015), SN 2012Z (Yamanaka et al. 2015), and SN 2014ck (Tomasella et al. 2016)
    are also plotted for comparison.}
    \label{fig:fig-4}
\end{figure}

In the following analysis, we further examine the explosion time and rise time of the light curve for SN 2014ek.
To better constrain the pre-maximum-light evolution of the light curves, the unfiltered data points from the TNTS are also
included in the analysis, as shown in Fig.~\ref{fig:fig-4}. As the unfiltered light curve is very similar to the $R$-band, we combine
them to determine the rising time and explosion date for SN 2014ek. Assuming that the luminosity evolution follows the
``expanding fireball" model (Riess et al. 1999), i.e., $f\propto (t-t_{0})^{n}$ (where "f" represents the flux, and t$_{0}$ denotes
the first-light time and it is also regarded as the explosion time in this paper), we derive $t_0$=$-14.82\pm0.95$
and n = 0.56$\pm$0.07 for SN 2014ek (see also the bold solid curve in Fig. 4). This corresponds to a $R$-band rising time of $14.82\pm0.95$ days
and an explosion date on MJD 56948.16$\pm$0.96. Fitting to the $R$-band light curve of SN 2012Z yields a similar estimate of the rise time, i.e., 14.06$\pm$1.69 days. The corresponding rise time is estimated as 13.46$\pm$0.38 days and 13.14$\pm$1.54 days for SN 2014ck and SN 2011ay, respectively, which is slightly shorter than that of SN 2014ek. In comparison, SN 2005hk seems to show an apparently
longer rise time of 16.65$\pm$0.25 days. Considering that the explosion time of each band is the same, the deduced rise time to the
maximum light of SN 2014ek ranges from $\sim10$ days in the $B$ band to $\sim17$ days in the $I$ band.

\begin{figure}
	\includegraphics[width=\columnwidth]{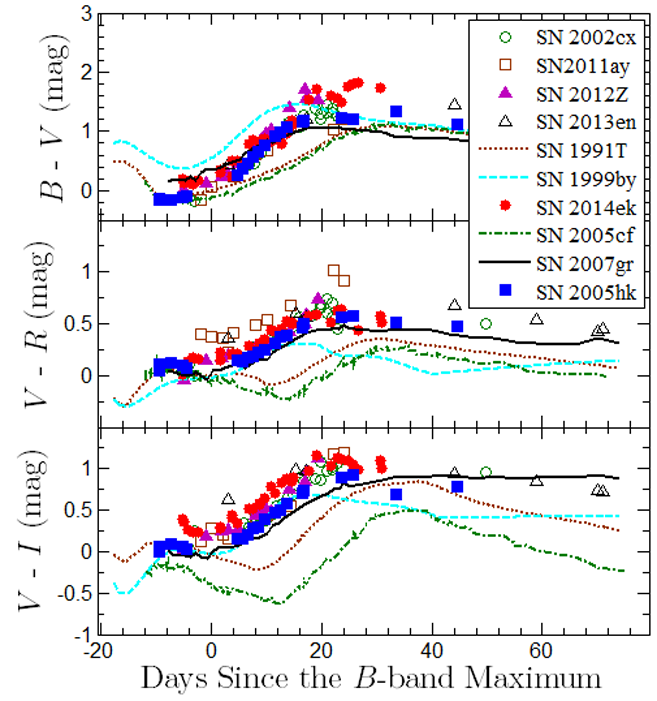}
    \caption{The $B-V$, $V-R$, and $V-I$ colors of SN 2014ek compared with those of SNe 2002cx (Li et al. 2003), 2005hk (Phillips et al. 2007), 2005cf (Wang et al. 2009), 2011ay (Szalai et al. 2015), 2012Z (Yamanaka et al. 2015), 2013en (Liu et al. 2015), 1991T (Lira et al. 1998), 1999by (Bonanos et al. 1999), and 2007gr (Chen et al. 2014). The color curves have been corrected for the Galactic reddening and the host-galaxy reddening whenever possible.}
    \label{fig:fig-5}
\end{figure}

\begin{figure}
	\includegraphics[width=\columnwidth]{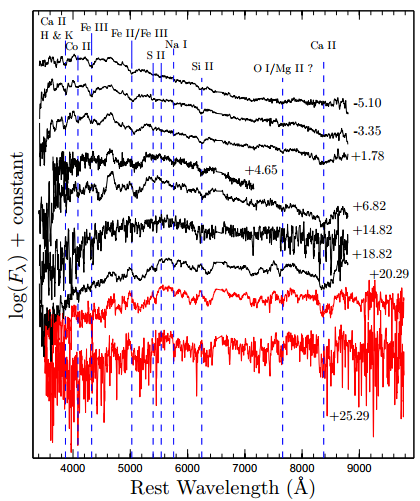}
    \caption{Optical spectral evolution of SN 2014ek. The spectra obtained with the Lijiang 2.4-m telescope are shown in upper region, while those obtained with the LCO FTN 2.0-m telescope are shown in the underside. All of the spectra have been smoothed with a bin of about 14\AA.}
    \label{fig:fig-6}
\end{figure}

\begin{figure}
	\includegraphics[width=\columnwidth]{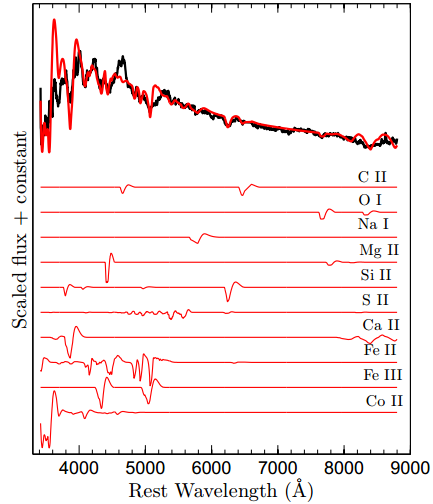}
    \caption{Optical spectrum of SN 2014ek taken at t$\sim$+1.8 days (black). The best-fit SYNAPPS synthetic spectrum is overplotted. The contribution of each ion is shown in the lower part of the plot.}
    \label{fig:fig-7}
\end{figure}

\begin{figure*}
	\includegraphics[width=17cm]{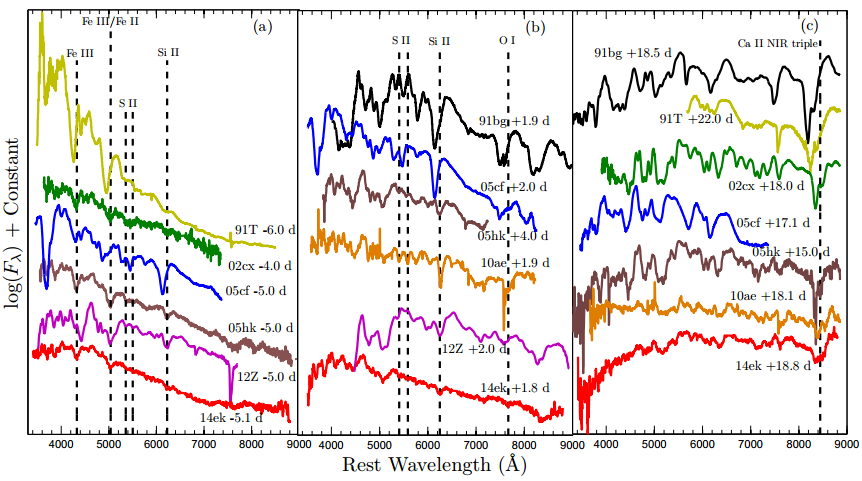}
    \caption{The spectra of SN 2014ek at t$\sim$$-$5.1 days, +1.8 days, and +18.8 days, compared with the spectra of SNe 2002cx (Li et al. 2003), 2005hk (Phillips et al. 2007), 2010ae (Stritzinger et al. 2014), 2012Z (Yamanaka et al. 2015), 2005cf (Wang et al. 2009), 1991T (Mazzali et al. 1995), and 1991bg (Turatto et al. 1996) taken at similar phases. All spectra shown here have been corrected for the reddening of the Milk Way and redshift of the host galaxy.}
    \label{fig:fig-8}
\end{figure*}

\subsection{Color Curves}

Fig.~\ref{fig:fig-5} shows the color evolution of SN 2014ek and other comparison SNe Iax. The color curves of different subclasses of SNe Ia are also overplotted. As it can be seen, SN 2014ek becomes progressively red after the first detection and it reaches the
reddest color at t$\approx$20-25 days after the maximum light. During the period from t$\sim$ +20 to +25 days, SN 2014ek seems to show
the reddest $B - V$ color among our comparison sample, which is consistent with the deficiency of the flux at shorter wavelength as
revealed by the spectra at similar phase (see Figure 6). After that, the color curves evolve bluewards, as similarly seen in other
comparison SNe Iax, but large scatter exists in the late-time evolution.

\subsection{Spectroscopy}

The spectral sequence of SN 2014ek is presented in Fig.~\ref{fig:fig-6}. These spectra are characteristic of typical SNe Iax, showing weak IMEs and dominant features of Fe {\sc ii}, Fe {\sc iii}, and Co {\sc ii} lines. All the spectra have been corrected for the recession velocity of the host galaxy (6933 km s$^{-1}$). The main lines are labeled in the spectra in light of the identifications obtained in previous studies of SN 2002cx-like object SN 2005hk (i.e., Phillips et al. 2007).

\subsubsection{Spectroscopic Evolution}

To better demonstrate the spectral features of SN 2014ek, we show in Fig.~\ref{fig:fig-7} the t = $+$1.8 day spectrum and the synthetic spectrum computed using the SYNAPPS code (Thomas et al. 2011). The SYNAPPS fit gives a black-body temperature of $T_{ph}$=6800 K
and an expansion velocity of $v_{ph}$=5200 km s$^{-1}$ for the photosphere. The comparison clearly favors for the presence of IMEs
(Si {\sc ii}, Ca {\sc ii}, S {\sc ii}, C {\sc ii}, and O {\sc i}), and IGEs such as Fe {\sc ii}, Fe {\sc iii}, and Co {\sc ii} in the
near-maximum-light spectrum. In particular, the prominent Fe {\sc ii}, Fe {\sc iii}, and Co {\sc ii} features are
reminiscent of the spectral features of SN 1991T-like subclass which has much higher photospheric temperature at similar phase,
distinguishing from those of the normal SNe Ia. There are still some minor absorptions that cannot be properly identified
for pre-maximum spectra, such as those at $\sim$6000\AA\ and 7000\AA\ etc., respectively, though the presence of these features may
be affected by lower signal-to-noise ratio of the spectrum. Those minor absorptions are also detected in the pre-maximum
spectra of some other SNe Iax such as SN 2011ay and SN 2007qd (McClelland et al. 2010; Szalai et al. 2015); and the absorptions
at $\sim$6000\AA\ can be identified as Fe {\sc ii}, while the absorption at $\sim$7000\AA\ might be attributed to O {\sc ii}
or C {\sc ii}.

In Fig.~\ref{fig:fig-8}, we compare the spectra of SN 2014ek with those of some well-observed SNe Iax at t$\sim$ -5 d, +2 d, and +18 d (relative to the $B$-band maximum), respectively. At t$\sim$5 days before the maximum light, SN 2014ek exhibits a blue continuum and weak absorptions of Fe {\sc ii}/Fe {\sc iii}, Si {\sc ii} 6355, and S {\sc ii} lines, similar to SN 2005hk (and perhaps SN 2002cx). While the W-shaped S {\sc ii} and O {\sc i} 7774 absorption features are clearly detected in SN 2005hk but they are almost invisible in SN 2014ek.
At this phase, SN 2012Z is found to have relatively broader and stronger line profiles for the above spectral features, in particular
the Fe {\sc ii}/{\sc iii} multiplet at $\sim$5000\AA, perhaps suggesting that it experienced an energetic explosion. At around the
maximum light, the Si {\sc ii} 6355, Ca {\sc ii}, and Fe {\sc ii}/{\sc iii} lines becomes stronger in SN 2014ek, and the S {\sc ii}
and O {\sc i} lines are visible in the spectrum. The O {\sc i} $\lambda7774$ absorption may be mixed with very weak Mg {\sc ii} line.
In comparison with SN 2014ek, SN 2010ae seems to have relatively deeper and stronger S {\sc ii} and Si {\sc ii} lines at
this phase. By t$\sim$2 weeks from the peak, the spectrum is dominated by some features of Fe {\sc ii} lines and Ca {\sc ii}
near-infrared (NIR) triplet. The relative strength of the three components of Ca {\sc ii} show obvious differences among the
comparison SNe Iax. No prominent features of He {\sc i} lines can be detected in the spectra of SN 2014ek.

\begin{figure}
	\includegraphics[width=\columnwidth]{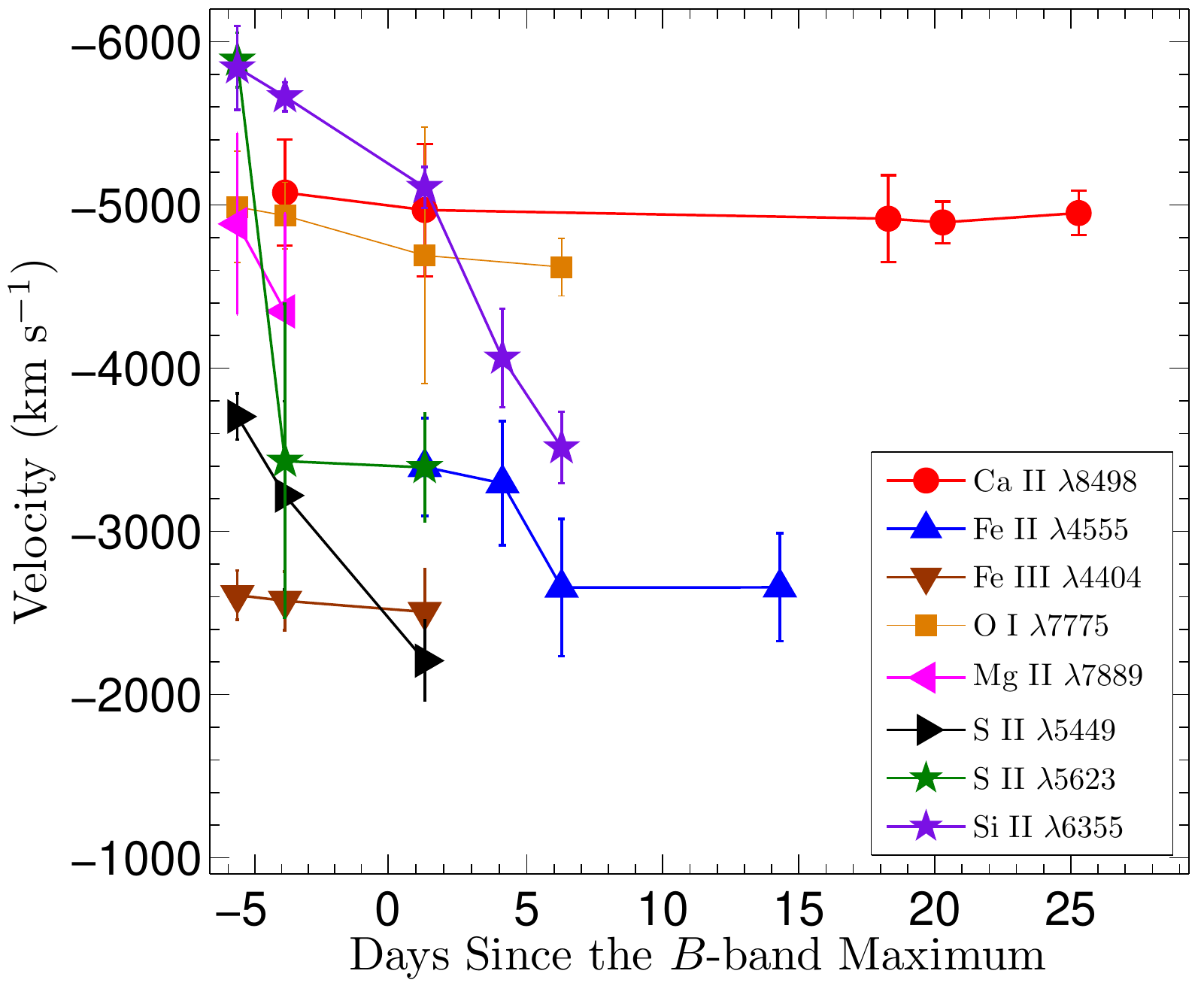}
    \caption{Velocity evolution of some absorption lines in the spectra of SN 2014ek near the $B$-band maximum light.}
    \label{fig:fig-9}
\end{figure}

\begin{figure}
	\includegraphics[width=\columnwidth]{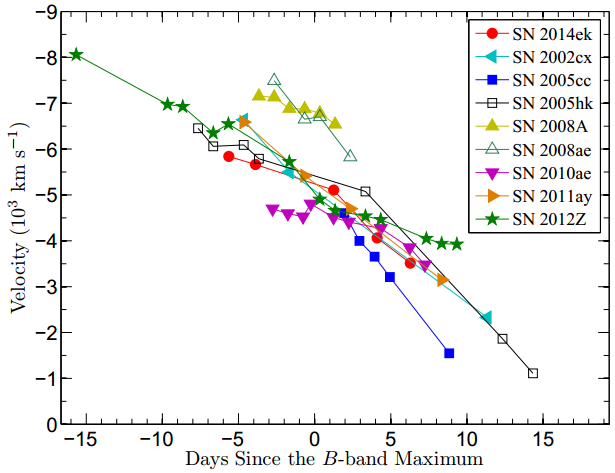}
    \caption{Comparison of velocity evolution of Si {\sc ii} $\lambda6355$ between SN 2014ek and some other SNe Iax. Spectra of SN 2005cc are from Blondin et al. (2012), while references for the spectra of other SNe Iax are listed in the last
    column of Table 5).} \label{fig:fig-10}
\end{figure}

\subsubsection{Ejecta Velocity}
In this subsection, we examined the ejecta velocity of SN 2014ek inferred from absorptions of different ions in the spectra.
Fig.~\ref{fig:fig-9} shows the velocities inferred from Si {\sc ii}, Ca {\sc ii}, O {\sc i}, and Fe {\sc ii} absorptions.
The above velocity measurements may have larger uncertainties due to lower S/N ratio of the spectra and line blending between
different ions. For example, the O {\sc i} $\lambda7773$ line may be mixed with weak Mg {\sc ii} line according to the SYNAPPS modeling.
Assuming one gaussian component in the fitting of Si {\sc ii} 6355 absorption, we got an expansion velocity of
$\sim$5000km s$^{-1}$ for SN 2014ek from the t$\sim$0 day spectrum, which is comparable to other SNe Iax such as SN 2002cx and
SN 2005hk at similar phases. For SNe Iax, however, the Si {\sc ii} 6355 absorption tends to blend with the neighboring
Fe {\sc ii} absorption (i.e., Fe {\sc ii} $\lambda6456$\AA) soon after the peak, and this will make the measurements of their Si II
velocities be biased towards smaller values. For example, the Si~II velocity of SN 2011ay was estimated as $\sim$5600 km s$^{-1}$
near the maximum light by assuming a single feature in the fit (Silverman et al. 2011; Foley et al. 2013), while a much higher
Si~II velocity (i.e., $\sim$9000km s$^{-1}$) can be determined if considering the contamination of the nearby iron lines
(Szalai et~al. 2015). For SN 2014ek, the near-maximum-light Si~II velocity can increase to $\sim$7200 km s$^{-1}$ when
taking into account the Fe II contamination.

The ejecta velocities derived for SN 2014ek from different ions are overall much lower than the corresponding values of
normal SNe Ia and SNe Ibc measured at similar phases. The Si II velocity tend to show a rapid decrease with time, with a gradient of
$195\pm34$ km s$^{-1}$ d$^{-1}$. In contrast, the velocity of Ca~II NIR triplet remains almost a constant during the period from one
week before the peak to three weeks after that, which may be due to that Ca {\sc ii} has a low excitation energy. And SN 2012Z shows similar evolution for the velocity of Ca~II NIR triplet (Yamanaka et al. 2015). To get a closer
inspection of the velocity evolution among different SNe Iax, we show in Fig.~\ref{fig:fig-10} the Si {\sc ii} $\lambda6355$ velocity evolution for SN 2014ek and other SNe Iax. It is readily seen that SN 2014ek has a velocity evolution similar to
SN 2005hk and SN 2002cx. Instead, SN 2005cc shows a faster velocity evolution with velocity gradient of 443$\pm$70; while SN 2008A and SN 2010ae have velocity gradients of 122$\pm$52 and 123$\pm$76, respectively.

\begin{figure}
	\includegraphics[width=\columnwidth]{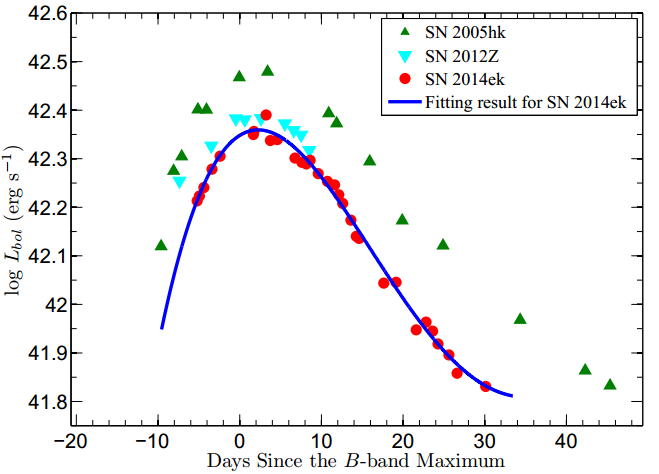}
    \caption{The pseudo-bolometric light curve of SN 2014ek. The polynomial fit to the observed data is shown as a line. For comparison, the pseudo-bolometric light curves of SN 2005hk (Phillips et al. 2007) and SN 2012Z (Yamanaka et al. 2015) are also plotted as triangles and inverted triangles, respectively.}
    \label{fig:fig-11}
\end{figure}

\begin{figure*}
	\includegraphics[width=17cm]{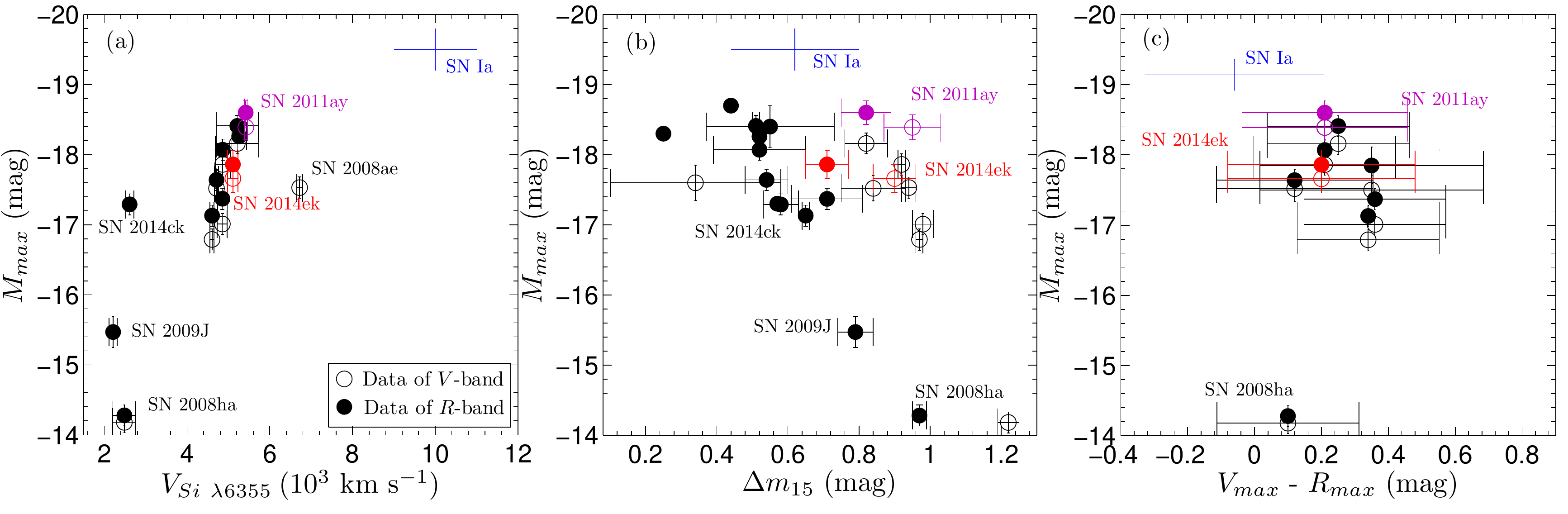}
    \caption{The absolute $VR-$band peak magnitudes plotted against the Si {\sc ii} $\lambda6355$ velocity measured around the maximum light, post-peak decline rate $\Delta{m_{15}}$, and peak $V - R$ color for SN 2014ek and other SNe Iax with available photometric and spectroscopic data. The cross represents the ranges of the corresponding parameters obtained for normal SNe Ia.}
    \label{fig:fig-12}
\end{figure*}

\begin{figure*}
	\includegraphics[width=17cm]{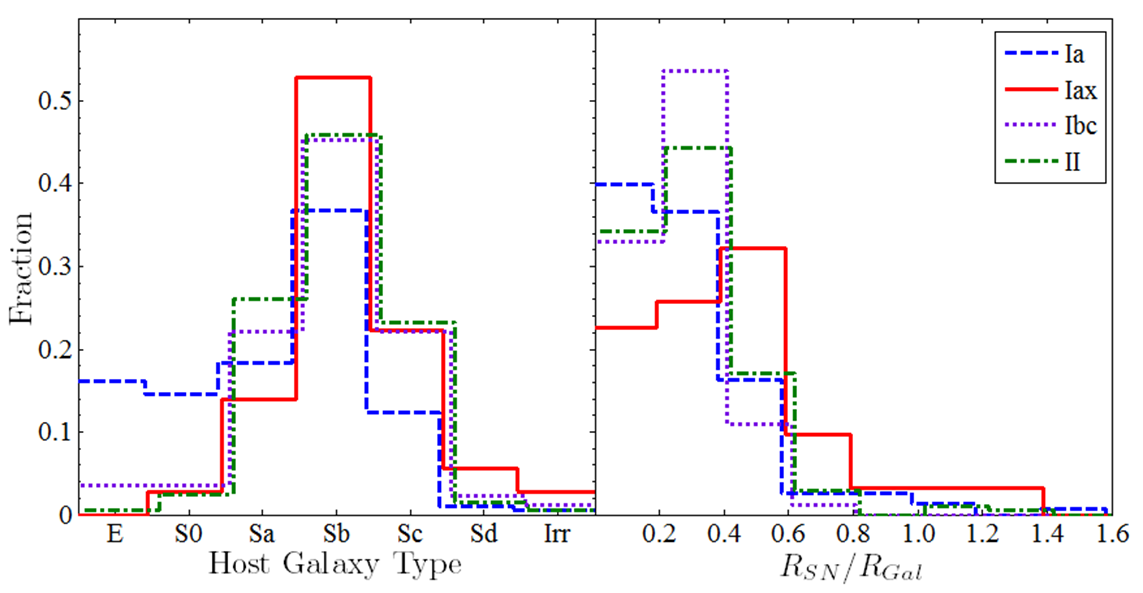}
    \caption{Left panel: Histogram distribution of host galaxy types for SNe Iax, SNe Ia, SNe Ibc and SNe II. Right panel: Histogram distribution of the projected distance of the supernovae from the galactic centre, normalized to the optical radius $R_{SN}/R_{gal}$, for SNe Iax, SNe Ia, SNe Ibc and SNe II.}
    \label{fig:fig-13}
\end{figure*}

\subsection{Bolometric Light Curve}
In this subsection, we constructed the bolometric light curve of SN 2014ek using the multicolor light curves presented in Section 3. As this peculiar SN is overall spectroscopically and photometrically similar to SN 2005hk in the optical band, we thus take the $BVRIJHK$-band light curve of the latter as a template to infer the flux contribution in the near-infrared (NIR) bands for the former. For SN 2005hk, the fractional contribution of the NIR flux to the bolometric flux is estimated to vary from 14.5\% at t=$-$5.2 days to 28.7\% at t= +31.0 days, based on the published data from Phillips et al. (2007). This flux ratio of NIR emission is then used to  build the pseudo-bolometric light curve of SN 2014ek, as shown in Fig.~\ref{fig:fig-11}. The bolometric light curves of SN 2005hk and SN 2012Z are overplotted for comparison (Phillips et al. 2007; Yamanaka et al. 2015). A polynomial fit to the light curve of SN 2014ek gives a peak bolometric luminosity as L$_{max}\approx$2.29$\times$10$^{42}$ erg s$^{-1}$, which seems to be fainter than SN 2005hk by a factor of $\sim1.3$.

With the derived bolometric luminosity, we can estimate the synthesized $^{56}$Ni mass. Assuming the Arnett's law (Arnett 1982) and the maximum luminosity can be produced by the radioactive $^{56}$Ni (Stritzinger \& Leibundgut 2005; Ganeshalingam et al. 2012), we estimate the mass of ${}^{56}$Ni produced during explosion as about $0.08 M_{\odot}$, consistent with estimates of other SNe Iax, which have a range from 0.03M$_{\odot}$ to 0.3 M$_{\odot}$. Different models have been proposed to explain the observed scatter of the nickel mass synthesized in the explosion of SNe Iax. One of these models is a hybrid C-O-Ne white dwarfs with a weak pure deflagration (Denissenkov et al. 2015), while the fall-back explosion of massive stars whose production may be a black hole or a compact star is also proposed (Moriya et al. 2010).

\begin{figure*}
	\includegraphics[width=17cm]{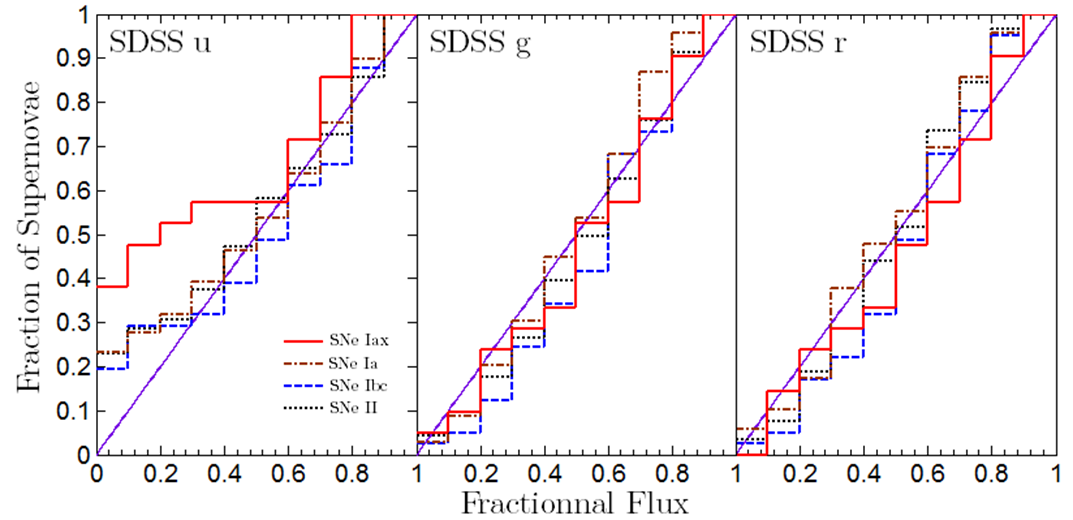}
    \caption{Histogram distribution of the fractional flux of the host-galaxy light at the location of SN progenitors in the SDSS $u$, $g$, and $r$ bands. The dash-dotted diagonal lines represent the case that the SN progenitors follow exactly the distribution of the galaxy light.}
    \label{fig:fig-14}
\end{figure*}

\section{Discussion}

\subsection{Luminosity of SNe Iax}

In this subsection, we examined the relations between absolute $VR-$band peak magnitudes and some photometric/spectroscopic parameters including post-maximum decline rate $\Delta m_{15}$, near-maximum-light Si {\sc ii} velocity, and peak $V - R$ color, based on an updated sample of SNe Iax (N=45). The relevant parameters of this SN sample and their host galaxies are listed in Table~\ref{tab:table 5}. Ejecta velocity inferred from Si {\sc ii} absorption line has been proposed to be an important parameter to distinguish different subclasses of SNe Ia (i.e., Wang et al. 2009, 2013). Fig.~\ref{fig:fig-12}(a) shows the absolute peak $RV$-band magnitudes as a function of Si {\sc ii} velocity measured around the maximum light. Most SNe Iax, including SN 2014ek, seem to have Si {\sc ii} velocities clustering in a narrow range of 4000-5000 km s$^{-1}$. For the SNe Iax within such a velocity range, their luminosity might be correlated with the Si~II velocity. Note that there exists a subgroup with slower Si {\sc ii} velocity at around 2000 km s$^{-1}$, i.e., SN 2008ha, SN 2009J, and 2014ck, which shows large scatter in their peak luminosity, ranging from $-$14.0 mag to $-$17.5 mag in the $R$ band. This complicates our understanding of the luminosity-velocity relation observed for SNe Iax. Fig.~\ref{fig:fig-12}(b) shows the absolute peak magnitudes M$_{V,R}$(max) plotted against $\Delta m_{15}$. SN 2014ek exhibits properties of a typical SN Iax like SN 2002cx, with a post-maximum decline rate of $\Delta m_{15}$ = 0.71$\pm$0.06 mag in $R$ and 0.90$\pm$0.06 mag in $V$, respectively. An anti-correlation seems to exist between peak magnitude and post-peak decline rate, but it does not hold for the rest sample when excluding SN 2008ha and SN 2009J which both have very low luminosity. Thus, more sample of SNe Iax with luminosity lying between the luminous and faint ends is needed to confirm or refute this trend. In Fig.~\ref{fig:fig-12}(c), we show the relation between the M$_{V,R}$(max) and the $V_{max} - R_{max}$ color. One can see that SNe Iax have on average redder colors than normal SNe Ia, but their peak luminosities do not show any significant correlation with the peak $V - R$ colors.

Compared to subluminous SN 1991bg-like subclass of SNe Ia, SNe Iax have similar peak luminosity but they show an obviously slower decline rate and lower ejecta velocity. One possible explanation, as proposed by Li et al. (2003), is that lower ejecta velocity results in a smaller escape possibility and a longer diffusion timescale for the $\gamma$-ray photons, which makes the photosphere cool down slowly and the SN have a slower post-maximum decline rate. However, the color evolution of SNe Iax may imply a rapid change of temperature and indicates that there may be some other reasons to explain the low luminosity and slow decline rate of SNe Iax. Nevertheless, the overall lower luminosity suggests a weak explosion for SNe Iax, i.e., with a small amount of ${}^{56}$Ni being produced only at the surface of white dwarfs (Foley et al. 2013).

\subsection{Host Galaxies}
Previous studies have shown that SNe Iax tend to occur in late-type galaxies (i.e., Foley et al. 2009), and SN 2014ek seems to also follow this tendency. The host galaxy of SN 2014ek shows a close resemblance to a Sb galaxy based on its morphology from the SDSS images. In order to explore the statistical properties of SN Iax host galaxies, we collect the host-galaxy parameters of 45 SNe Iax from the literature and our own database. These parameters are presented in Table~\ref{tab:table 5}, including host-galaxy morphology, redshift (z$_{Helio}$), radial distribution (R$_{SN}$/R$_{gal}$), fractional flux in $ugr$ bands ($f_u$,$f_g$ and $f_{r}$) and metallicity of the supernova sites. Some photometric parameters of the SNe are also listed, which includes absolute $V$-band peak magnitudes, $\Delta{m_{15}}$ in the $R$ band, and ejecta velocity of Si {\sc ii} $\lambda6355$ near the $B$-band maximum light.
For the 37 SNe Iax with hubble types determined or judged for their host galaxies, 29 (with a fraction of 78.3\%) are found to occur
in Sb-Irr galaxies, 7 in Sa-Sba galaxy (SN 2002bp, SN 2006hn, SN 2008A, SN 2011ay, SN 2011ce, SN 2013en, and SN 2014cr), and
1 in S0 galaxy (SN 2008ge), while none was found in elliptical galaxy. This indicates that SNe Iax are skewed to the late-type
spiral galaxies like core-collapse SNe (i.e., Li et al. 2011), favoring that they should arise from younger progenitor populations. The histogram distributions of host galaxy type for SNe Ia, SNe Iax, SNe Ibc and SNe II are shown in the left panel of Fig. 13. Samples of SNe Ia, SNe Ibc and SNe II in Fig. 13 are taken from Wang et al. (2013). In the right panel of Fig. 13, we further examine the radial distribution of SNe Iax and other subclasses of SNe, which might be related to the age and/or metallicity of the supernova sites. This radial distance from the galactic centre, normalized to the optical radius of the host galaxy, RSN/Rgal, is available for 31 of our sample of SNe Iax. We found that nearly 43 percent of them are located in regions with 0.4 < $R_{SN}/R_{gal}$ < 0.8, while this fraction is only about 11 percent for SNe Ibc and 20 percent for SNe II, respectively. This contrast in radial distribution indicates that SNe Iax tend to occur in outer regions of the galaxy and their progenitors may have relatively lower metallicity, but more sample of SNe Iax are needed
to confirm this trend.

There are a few measurements of oxygen abundance for the host galaxies of SNe Iax, which are in a range of 12+log(O/H)$=$8.16$\sim$8.87 (Magee et al. 2017; Lyman et al. 2018), somehow lower than the solar abundance (Caffau et al. 2008). This indicates that the site of SNe Iax may have relatively lower metallicity compared to other subtypes of SNe Ia, but comparable to that of SNe II or SNe Ibc. Moreover, we examine the radial distribution of SNe Iax, which might be also related to the age and/or metallicity of the supernova sites. This radial distance from the galactic center, normalized to the optical radius of the host galaxy, R$_{SN}$/R$_{gal}$ is available for 31 of our sample. We found that nearly half of them are found to locate in regions with 0.3$<$R$_{SN}$/R$_{gal}$$<$0.6, similar to the radial distribution of core-collapse SNe, especially the SNe Ibc (Li et al. 2011). Note that Lyman et al. (2018) recently carried out an analysis of the metallicity for the host galaxies and explosion sites of large sample of SNe Iax through both integral-field and long-slit spectroscopy, and they also found that SN Iax explosion site metallicity distribution is similar to that of core-collapse SNe. This result is consistent with our analysis from the above radial distribution and the fractional flux at the SN site discussed below.

We also measure the fractional flux of the SN site, which is the fraction that the total host light in pixel is fainter than or equal to the light in the pixel at the location of the SN. This parameter can be used to trace the difference/similarity between luminosity at SN locations and light of the hosts (Fruchter et al. 2006). As the fractional flux is independent of galaxy morphology and size, it is thus a better parameter to quantify the correlation between the SN progenitor environments and the light of their hosts (i.e., Wang et al. 2013). Such an analysis can be done for 21 SNe Iax of our sample by using the host-galaxy images from SDSS DR12 (Alam et al. 2015) and Panstarrs (Chambers et al. 2017), and the results are also tabulated in columns (9)-(11) of Table~\ref{tab:table 5}. The distribution of the fractional fluxes of our SN Iax sample in the $ugr$ bands is shown in Fig.~\ref{fig:fig-14}. The fractional flux distributions of SNe Ia, SNe Ibc, and SNe II are overplotted for comparison, and the measurements of these samples are taken from Wang et al. (2013). As one can see that the progenitor populations of SNe Iax belong to the brighter regions of the host galaxies, which is similar to that of SNe Ibc (especially in the $g$ and $r$ bands), suggesting that SNe Iax may arise in larger star-forming regions that produce more-massive stars. This is also evidenced by the K-S test results, as tabulated in Table~\ref{tab:table 6}. One can see that SNe Iax and SNe Ia trace the host-galaxy light differently, with a very low probability (i.e., P = 0.04 in $r$) that they come from the same stellar populations. SNe Iax also show some differences from SNe II in the light distribution of progenitor populations, with a KS test result of $P = 0.13$ in $g$ and P = 0.17 in $r$, respectively. In the $u$ band, however, we notice that SNe Iax seems to show large differences from all kinds of SNe in the fractional flux distribution. This is likely related with the relatively larger uncertainty in the SDSS$-u$ band photometry. As for the comparison between SNe Iax and SNe Ibc, a P value of 0.33 obtained in $g$ band and 0.37 in $r$ band indicates that the hypothesis that they share a similar fractional flux (and hence environment) distribution cannot be rejected.

\subsection{Progenitors and Explosion Models}

For the 45 objects that are classified as SNe Iax, their ejecta velocities, strength of absorption lines, and peak luminosity show large differences. Most of the sample are found to be similar with SN 2002cx, such as SN 2014ek presented in this paper. There is also a small fraction of SNe Iax which show close resemblance to SN 2008ha. SN 2008ge may belong to another subtype which shows very broad light curve relative to SN 2002cx (Foley et al. 2010).

Previous studies indicate that the progenitors of SNe Iax may be a WD with a companion star exploding as weak deflagration (Jordan et al. 2012; Kromer et al. 2013; Fink et al. 2014), consistent with the S {\sc ii} and O {\sc i} features in their spectra, but it can not explain properly the properties of the faint members like SN 2008ha which was even argued to result from fallback explosion of massive progenitors (Valenti et al. 2009; Kromer et al. 2015). Moreover, Si {\sc ii} and S {\sc ii} lines are not unique of thermonuclear explosions, which are often seen in the spectra of some SNe Ibc (Valenti et al. 2008; Brown et al. 2007).

On the other hand, analysis of the Hubble Space Telescope (HST) images prior to the explosion of SN 2012Z shows that its progenitor system is consistent with a blue star, perhaps consisting of a WD and a helium companion (McCully et al. 2014). The pre-explosion HST images suggest that the progenitor SN 2008ge should arise from thermonuclear explosion of a WD star (Foley et al. 2010). However, it should be noted that SN 2008ge may not represent a typical SN Iax, which exploded in a lenticular galaxy with a stellar-population age of 9.5 Gyr. Owing to that its first spectrum was taken at t$\sim$ 40 days after the peak, the exact spectroscopic classification of this object is still uncertain. Moreover, the very broad light curve indicates it is different from either the SN 2002cx-like SNe Iax or the SN 2008ha-like SNe Iax.

Given the existing subtypes like SN 2002cx, SN 2008ha, and perhaps SN 2008ge, it is possible that SNe Iax may have multiple progenitor systems, although the so-called hybrid C-O-Ne WDs progenitor scenario may be able to explain the extremely low ${}^{56}$Ni mass of faint ones of SNe Iax (Denissenkov et al. 2015).

\section{Conclusion}
We present the optical photometry and spectra of a 2002cx-like supernova SN 2014ek in this paper. SN 2014ek reached its peak magnitude m$_{V}$=17.50$\pm$0.04 mag on 2014 Oct. 30.89 UT, with a decline rate of $\Delta$m$_{15}$($V$)=0.90$\pm$0.06 mag. Combining the early-time light curve with the pre-discovery detection limit, we derive a rising time of $\sim$14.8 days in the $R$ band for SN 2014ek, much faster than a typical SN~Ia. At around the maximum light, the Si~II velocity is found to be about 5100 km s$^{-1}$ and the absolute magnitude is estimated as $-$17.66$\pm$0.20 mag in $V$, which are consistent with those of SN 2002cx and SN 2005hk. Our discovery and followup observations of SN 2014ek help further populate the group between the over-luminous ones and faint ones of SNe Iax. This peculiar SN also shows similar spectral features and evolution with SN 2002cx and SN 2005hk, but the line strengths in the spectra of SN 2014ek seems to be relatively weaker at late time. We constructed the optical near-infrared pseudo-bolometric light curve of SN 2014ek and found that its peak luminosity as $(log\ L)_{max}\approx42.36\ erg\ s^{-1}$, corresponding to a ${}^{56}$Ni mass of $0.08M_{\odot}$ synthesized in the explosion.

By collecting a large sample of SNe Iax (N=45) from the literature and our own database, we explored the observed properties of SNe and their host galaxies. We found that the peak luminosity of these SNe Iax might have a loose correlation with their ejecta velocity, decline rate, or peak color, but these weak correlations depend largely on two peculiar events like SN 2008ha and SN 2009J in the sample. Furthermore, we found that the hosts of SNe Iax are highly skewed to the late-type spiral galaxies, with more than 78\% of the sample coming from Sb or later than Sb-type galaxies and only one in lenticular galaxy. The distribution of fractional fluxes of the light at SN positions is also explored, and SNe Iax tend to be associated with bright, star-forming regions within their host galaxies, which are more similar to that of SNe Ibc rather SNe Ia and even SNe II. This results suggest that SNe Iax, or at least part of them, may arise from young, massive progenitor systems. In the future, it will be interesting to examine whether SN 2008ha-like SNe can be regarded as a member of SN 2002cx-like explosions extending to the faint end or they represent stellar explosions with distinct physical mechanisms or progenitor models when a larger sample of similar properties are available.

\section*{Acknowledgements}
We acknowledge the support of the staff of the Lijiang 2.4m and Xinglong 2.16m telescope. Funding for the LJT has been provided by Chinese Academy of Sciences and the People's Government of Yunnan Province. The LJT is jointly operated and administrated by Yunnan Observatories and Center for Astronomical Mega-Science, CAS. This work is supported by the National Natural Science Foundation of China (NSFC grants 11178003, 11325313, and 11633002), and the National Program on Key Research and Development Project (grant no. 2016YFA0400803). J.-J. Zhang is supported by the National Science Foundation of China (NSFC, grants 11403096, 11773067), the Youth Innovation Promotion Association of the CAS, the Western Light Youth Project, and the Key Research Program of the CAS (Grant NO. KJZD-EW-M06). T.-M. Zhang is supported by the NSFC (grants 11203034). This work was also partially Supported by the Open Project Program of the Key Laboratory of Optical Astronomy, National Astronomical Observatories, Chinese Academy of Sciences. This work makes use of observations from Las Cumbres Observatory. DAH, CM, and GH are supported by the US National Science Foundation grant 1313484. Support for IA was provided by NASA through the Einstein Fellowship Program, grant PF6-170148.

\begin{table*}
	\centering
    \caption{Photometric Standards in the SN 2014ek Field}
    \label{tab:table 1}
	\begin{tabular}{ccccccccccccc}
		\hline
		 & $RA$ & $DEC$ & $U$ & $B$ & $V$ & $R$ & $I$ & $u$ & $g$ & $r$ & $i$\\
         & & & (mag$^a$) & (mag) & (mag) & (mag) & (mag) & (mag) & (mag) & (mag) & (mag)\\
		\hline
		1& $23^h55^m57.58^s$ & +$29^\circ21'43.89''$ & 16.94(03) & 17.04(01) & 15.41(01) & 14.52(01) & 13.63(01) & 17.37(01) & 16.32(01)& 14.73(01)& 14.12(01)\\
		2& 23:56:16.73 & +29:21:52.78 & 17.66(01) & 16.53(03) & 15.58(01) & 14.53(01) & 14.07(01) & 16.53(01) & 16.78(01)& 14.68(01)& 14.49(01)\\
		3& 23:55:57.53 & +29:21:17.25 & 17.50(01) & 17.50(03) & 16.89(01) & 16.58(01) & 16.13(01) & 18.33(01) & 17.15(01)& 16.72(01)& 16.55(01)\\
		4& 23:56:18.21 & +29:21:02.31 & 18.12(03) & 18.24(01) & 17.63(01) & 17.31(01) & 16.85(01) & 18.92(02) & 17.89(01)& 17.46(01)& 17.27(01)\\
		5& 23:55:53.63 & +29:21:19.73 & 18.90(03) & 18.38(01) & 17.55(01) & 17.15(01) & 16.71(01) & 19.79(03) & 17.95(01)& 17.27(01)& 17.23(01)\\
		6& 23:55:50.63 & +29:21:03.1 & 18.71(03) & 18.08(01) & 17.15(01) & 16.66(01) & 16.09(01) & 19.59(03) & 17.61(01)& 16.82(01)& 16.54(01)\\
		\hline
        \multicolumn{9}{l}{$^a$ Note: Uncertainties, in units of 0.01 mag.}
	\end{tabular}
\end{table*}

\begin{table*}
	\centering
	\caption{Optical Photometry of SN 2014ek}
	\label{tab:table 2}
	\begin{tabular}{cccccccccccc}
		\hline
		MJD & Phase$^a$ & $U$ & $B$ & $V$ & $R$ & $I$ & $g$ & $r$ & $i$ & Unfiltered & Telescope\\
		 & (days) & (mag) & (mag) & (mag) & (mag) & (mag) & (mag) & (mag) & (mag) & (mag) & \\
		\hline
		56949.00 & -9.39 & ... & ... & ... & ... & ... & ... & ... & ... & 18.48(44) & TNTS \\
56952.00 & -6.39 & ... & ... & ... & ... & ... & ... & ... & ... & 17.88(41) & TNTS \\
56953.18 & -5.21 & ... & 18.17(02) & 17.93(04) & 17.89(02) & 17.46(02) & 17.95(03) & 17.98(03) & 17.74(04) & ... & LCO \\
56953.45 & -4.94 & ... & 18.12(04) & 17.95(03) & 17.82(03) & 17.51(04) & 17.92(05) & 17.91(03) & 17.78(06) & ... & LCO \\
56954.04 & -4.35 & 17.44(03) & 18.07(01) & 17.86(01) & 17.68(05) & 17.52(03) & ... & ... & ... & ... & LJT \\
56954.99 & -3.40 & 17.55(07) & 17.95(01) & 17.79(01) & 17.59(05) & 17.49(03) & ... & ... & ... & ... & LJT \\
56955.99 & -2.40 & 17.46(07) & 17.92(01) & 17.69(01) & 17.50(05) & 17.38(03) & ... & ... & ... & ... & LJT \\
56956.00 & -2.39 & ... & ... & ... & ... & ... & ... & ... & ... & 17.05(64) & TNT \\
56958.00 & -0.39 & ... & ... & ... & ... & ... & ... & ... & ... & 17.33(47) & TNT \\
56960.06 & 1.67 & ... & 17.94(02) & 17.55(02) & 17.28(04)& 17.19(04) & 17.70(03) & 17.37(05) & 17.66(08) & ... & LCO \\
56960.13 & 1.74 & 17.68(16) & 17.89(02) & 17.52(02) & 17.34(05) & 17.15(03) & ... & ... & ... & ... & LJT \\
56962.14 & 3.75 & ... & 18.10(03) & 17.55(03) & 17.32(05)& 17.03(05) & 17.74(04) & 17.51(03) & 17.49(04) & ... & LCO \\
56962.17 & 3.78 & 17.97(05) & 17.81(01) & 17.51(01) & 17.26(01)& 17.02(01)& ... & ... & ... & ... & LJT \\
56962.99 & 4.60 & ... & 18.08(02) & 17.53(02) & 17.28(05) & 17.10(03) & ... & ... & ... & ... & LJT \\
56965.16 & 6.77 & 18.52(05) & 18.46(01) & 17.61(02) & 17.29(05) & 17.02(03) & ... & ... & ... & ... & LJT \\
56966.00 & 7.61 & ... & 18.49(09) & 17.59(06) & 17.35(05)& 17.00(05)& ... & ... & ... & 17.40(31) & TNT \\
56967.00 & 8.61 & ... & 18.53(05) & 17.65(03) & 17.30(03)& 17.01(08)& ... & ... & ... & 17.64(30) & TNT \\
56967.08 & 8.69 & ... & 18.52(06) & 17.66(03) & 17.34(02)& 17.01(01)& ... & ... & ... & ... & LJT \\
56968.00 & 9.61 & ... & 18.67(08) & 17.75(04) & 17.36(04)& 17.02(05)& ... & ... & ... & 17.80(54) & TNT \\
56969.12 & 10.73 & ... & 18.76(06) & 17.87(06) & 17.40(05)& 17.01(09) & 18.28(07) & 17.65(05) & 17.53(05) & ... & LCO \\
56970.00 & 11.61 & ... & 18.83(05) & 17.84(02) & 17.39(02)& 17.07(02)& ... & ... & ... & 17.44(57) & TNT \\
56971.00 & 12.61 & ... & 18.90(06) & 18.04(03) & 17.55(02)& 17.14(03)& ... & ... & ... & 17.60(56) & TNT \\
56971.04 & 12.65 & ... & 18.83(02) & 17.99(01) & 17.51(01)& 17.07(01)& ... & ... & ... & ... & LJT \\
56972.00 & 13.61 & ... & 19.16(04) & 18.14(02) & 17.57(02)& 17.19(02)& ... & ... & ... & ... & TNT \\
56973.00 & 14.61 & ... & 19.41(09) & 18.18(02) & 17.63(02)& 17.25(03)& ... & ... & ... & ... & TNT \\
56973.21 & 14.82 & ... & 19.35(12) & 18.19(08) & 17.66(02)& 17.22(02)& ... & ... & ... & ... & LJT \\
56976.03 & 17.64 & ... & 20.03(05) & 18.43(02) & 17.82(06) & 17.38(03) & ... & ... & ... & ... & LJT \\
56977.53 & 19.14 & ... & 20.32(06) & 18.54(02) & 17.92(01)& 17.31(01)& ... & ... & ... & ... & LJT \\
56980.00 & 21.61 & ... & 20.37(11) & 18.72(04) & 18.18(03)& 17.61(05)& ... & ... & ... & ... & TNT \\
56981.21 & 22.82 & ... & 20.45(18) & 18.84(06) & 18.18(04)& 17.63(07) & 19.76(09) & 18.78(04) & 18.44(06) & ... & LCO \\
56982.00 & 23.61 & ... & 20.44(08) & 18.89(03) & 18.22(02)& 17.71(03)& ... & ... & ... & 18.17(38) & TNT \\
56983.17 & 24.78 & ... & 20.71(10) & 18.90(03) & 18.30(02)& 17.72(02)& ... & ... & ... & ... & LJT \\
56984.00 & 25.61 & ... & 20.79(11) & 18.93(04) & 18.32(03)& 17.81(06)& ... & ... & ... & 18.54(63) & TNT \\
56985.00 & 26.61 & ... & 20.83(08) & 18.96(03) & 18.49(03)& 17.89(05)& ... & ... & ... & ... & TNT \\
56989.04 & 30.65 & ... & 20.92(10) & 19.14(03) & 18.52(02)& 17.96(02)& ... & ... & ... & ... & LJT \\
56989.26 & 30.87 & ... & ... & 19.14(08) & 18.59(05) & 18.05(10) & 20.37(09) & 19.57(08) & 19.19(06) & ... & LCO \\
		\hline
        \multicolumn{12}{l}{$^a$ Relative to the $B$-band maximum (MJD = 56958.39).}
	\end{tabular}
\end{table*}

\begin{table*}
	\centering
	\caption{Spectroscopic Observation of SN 2014ek}
	\label{tab:table 3}
	\begin{tabular}{ccccccc}
		\hline
		UT Date & MJD & Phase$^a$ (days) & Exp.(s) & Telescope + Instrument & Range(\AA) & Standard Star\\
		\hline
		2014 Oct. 22 & 56953.29 & -5.10 & 2400 & YNAO 2.4m+YFOSC & 3500-9100 & Hilt600\\
2014 Oct. 24 & 56955.04 & -3.35 & 2400 & YNAO 2.4m+YFOSC & 3500-9100 & BD+28d4211\\
2014 Oct. 29 & 56960.17 & 1.78 & 2400 & YNAO 2.4m+YFOSC & 3500-9100 & BD+28d4211\\
2014 Nov. 1 & 56963.04 & 4.65 & 2700 & YNAO 2.4m+YFOSC & 3600-7400 & BD+28d4211\\
2014 Nov. 3 & 56965.21 & 6.82 & 2400 & YNAO 2.4m+YFOSC & 3500-9100 & Feige15\\
2014 Nov. 11 & 56973.21 & 14.82 & 2294 & YNAO 2.4m+YFOSC & 3500-9100 & G191B2B\\
2014 Nov. 15 & 56977.21 & 18.82 & 2700 & YNAO 2.4m+YFOSC & 3500-9100 & Feige110\\
2014 Nov. 17 & 56978.68 & 20.29 & 3600 & FTN 2m+FLOYDS & 3200-10000 & Feige34\\
2014 Nov. 22 & 56983.68 & 25.29 & 3600 & FTN 2m+FLOYDS & 3200-10000 & Feige34\\
		\hline
        \multicolumn{7}{l}{$^a$ Relative to the $B$-band maximum (MJD = 56958.39).}
	\end{tabular}
\end{table*}

\begin{table*}
	\centering
	\caption{Photometric Parameters of SN 2014ek}
	\label{tab:table 4}
	\begin{tabular}{ccccc}
		\hline
		Parameters$^a$ & $B$ & $V$ & $R$ & $I$ \\
		\hline
		$t_{max}$ (MJD) & $56958.39\pm0.34$ & $56960.89\pm0.18$ & $56962.98\pm0.13$ & $56964.86\pm0.17$ \\
Peak apparent magnitude (mag) & $17.89\pm0.12$ & $17.50\pm0.04$ & $17.26\pm0.04$ & $17.02\pm0.08$ \\
Peak absolute magnitude$^b$ (mag) & $-17.32\pm0.23$ & $-17.66\pm0.20$ & $-17.86\pm0.20$ & $-18.06\pm0.22$ \\
$\Delta{m_{15}}$ (mag) & $1.54\pm0.17$ & $0.90\pm0.06$ & $0.71\pm0.06$ & $0.52\pm0.11$ \\
		\hline
        \multicolumn{5}{l}{$^a$ We take the $34.99\pm0.20$ mag as distance modulus from NED to calculate the absolute magnitude.} \\
        \multicolumn{5}{l}{$^b$ Absolute magnitudes have been corrected for the extinctions of the Milkyway.}
	\end{tabular}
\end{table*}

\begin{table*}
	\centering
    \caption{Host Galaxy of SNe Iax Samples}
	\label{tab:table 5}
    \begin{sideways}
	\begin{tabular}{ccccccccccccc}
		\hline
		SN & Galaxy & Galaxy & $z_{Helio}$ & {M$_{R,peak}$} & $\Delta$m$_{15}$(R) & $v^0_{Si\ {\sc ii}}$ & {R$_{SN}$/R$_{gal}$}$^a$ & {$f_u$}$^b$ & $f_g$ & $f_r$ & 12+log(O/H) & Reference$^c$ \\
  &   & Type &   & (mag) & (mag) & ($10^3$ km s$^{-1}$) &   &   &   &   &   &  \\
		\hline
		SN1991bj & IC 344 & Sb & 0.018 & ... & ... & ... & 0.59 & ... & ... & ... & 8.37(01) & 2,3,16 \\
SN1999ax & SCP J140358.12+155106.9 & ... & 0.050 & ... & ... & ... & 0.56 & 0.17 & 0.44 & 0.44 & ... & 3 \\
SN2002bp & UGC 6332 & SBa & 0.021 & ... & ... & ... & 0.93 & 0.17 & 0.28 & 0.38 &  8.77(25) & 3,16 \\
SN2002cx & CGCG 44-035 & Sb & 0.024 & -17.64(15) & 0.54(06) & 4.71(05) & 1.23 & 0.30 & 0.60 & 0.63 & 8.36(01) & 2,3,4,16 \\
SN2003gq & NGC 7407 & Sbc & 0.021 & -17.37(15) & 0.71(10) & 4.85(12) & 0.10 & 0.75 & 0.92 & 0.92 & 8.33(02) & 2,3,16 \\
SN2004cs & UGC 11001 & Sc & 0.014 & ... & ... & ... & 0.47 & 0.88 & 0.88 & 0.82 & 8.43(01) & 3,16 \\
SN2004gw & PGC 16812 & Sbc & 0.017 & ... & ... & ... & 1.13 & ... & ... & ... & ... & 2,3 \\
SN2005P & NGC 5468 & Scd & 0.009 & ... & ... & ... & 0.44 & ... & ... & ... & 8.36(06) & 2,3,16 \\
SN2005cc & NGC 5383 & Sb & 0.008 & -17.13(15) & 0.65(01) & 4.60(05) & 0.05 & 0.99 & 1.00 & 1.00 & 8.50(02) & 2,3,16 \\
SN2005hk & UGC 272 & Sd & 0.013 & -18.07(15) & 0.52(13) & 4.86(18) & 0.45 & 0.24 & 0.31 & 0.26 & 8.34(09) & 2,3,5,16 \\
SN2006hn & UGC 6154 & Sa & 0.017 & ... & ... & ... & 0.25 & 0.85 & 0.91 & 0.92 & 8.50(02) & 2,3,16 \\
SN2007J & UGC 1778 & Sd & 0.017 & ... & ... & ... & 0.45 & ... & ... & ... & 8.38(02) & 2,3,16 \\
SN2007ie & SDSS J21736.67+003647.6 & ... & 0.093 & ... & ... & ... & ... & 0.00 & 0.67 & 0.75 & ... & 17 \\
SN2007qd & SDSS J020932.74-005959.6 & Sc & 0.043 & ... & ... & 2.80(10) & ... & 0.00 & 0.16 & 0.22 & 8.79(03) & 2,3,6 \\
SN2008A & NGC 634 & Sa & 0.016 & -18.41(15) & 0.51(01) & 5.22(51) & 0.38 & ... & ... & ... & ... & 2,3 \\
SN2008ae & IC 577 & Sc & 0.030 & ... & ... & 6.72(07) & 0.71 & 0.47 & 0.63 & 0.67 & 8.50(02) & 2,3,16 \\
SN2008ge & NGC 1527 & S0 & 0.004 & ... & ... & ... & 0.05 & ... & ... & ... & ... & 2,3,7 \\
SN2008ha & UGC 12682 & Irr & 0.005 & -14.28(15) & 0.97(02) & 2.48(28) & 0.31 & 0.27 & 0.64 & 0.69 & 8.16(15) & 2,3 \\
PTF09ego & SDSS J172625.23+625821.4 & ... & 0.104 & -18.30 & 0.25 & ... & ... & ... & ... & ... & ... & 17,19,20 \\
PTF09eoi & SDSS J232412.96+124646.6	& ... & 0.042 & -16.90 & ... & ... & ... & 0.15 & 0.79 & 0.78 & ... & 19,20 \\
SN2009J & IC 2160 & Sbc & 0.016 & -15.47(22) & 0.79(05) & 2.21(10) & 0.44 & ... & ... & ... & 8.41(02) & 2,3,16 \\
PTF10xk & ... & ... & ... & -17.30 & 0.57 & ... & ... & ... & ... & ... & ... & 19,20 \\
PTF10ujn$^d$ & ... & ... & 0.12 & ... & ... & ... & ... & ... & ... & ... & ... & 16 \\
SN2010ae & ESO 162-17 & Sb & 0.004 & ... & ... & 4.52(03) & 0.20 & ... & ... & ... & 8.40(18) & 3,8 \\
SN2010el & NGC 1566 & Sbc & 0.005 & ... & ... & ... & 0.10 & ... & ... & ... & 8.51(02) & 3,9,16 \\
SN2011ay & NGC 2315 & Sa & 0.021 & -18.60(17) & 0.82(07) & 5.42(04) & 0.24 & ... & ... & ... & ... & 3,9 \\
SN2011ce & NGC 6708 & Sa & 0.009 & ... & ... & ... & 0.16 & ... & ... & ... & ... & 3 \\
PTF11hyh & SDSS J014550.57+143501.9 & ... & ... & -18.70 & 0.44 & ... & ... & ... & ... & ... & ... & 19,20 \\
SN2012Z & NGC 1309 & Sbc & 0.007 & -17.92(02) & 0.52(01) & 5.26(18) & 0.69 & 0.11 & 0.31 & 0.24 & 8.51(31) & 3,10 \\
PS1-12bwh & CGCG 205-21 & Sbc & 0.023 & ... & ... & ... & 0.40 & 0.71 & 0.83 & 0.84 & 8.87(19) & 11,18 \\
iPTF13an & 2MASX J12141590+1532096 & Sb & 0.080 & ... & ... & ... & ... & 0.00 & 0.31 & 0.59 & ... & 19,20 \\
SN2013dh & NGC 5936 & Sb & 0.013 & ... & ... & ... & ... & 0.96 & 1.00 & 1.00 & ... & 20,21 \\
SN2013en & UGC 11369 & Sba & 0.015 & -18.40(30) & 0.55(18) & ... & 0.55 & ... & ... & ... & 8.46(03) & 12,16 \\
SN2014ck & UGC 12182 & Sbc & 0.005 & -17.29(15) & 0.58(05) & 2.61(10) & 0.13 & ... & ... & ... & ... & 13 \\
SN2014cr & NGC 6806 & Sa & 0.019 & ... & ... & ... & 0.30 & ... & ... & ... & ... & 17 \\
SN2014dt & NGC 4303 & Sc & 0.005 & ... & ... & ... & 0.31 & 0.84 & 0.87 & 0.92 & 8.52(02) & 16,17 \\
SN2014ek & UGC 12850 & Sc & 0.023 & -17.86(20) & 0.71(06) & 5.10(03) & 0.24 & 0.93 & 0.95 & 0.97 & 8.50(02) & 1,16 \\
SN2014ey & 2MASX J15042974+0219591 & Sbc & 0.031 & ... & ... & ... & 0.00 & ... & ... & ... & 8.43(02) & 15 \\
SN2015ce & UGC 12156 & Sbc & 0.017 & ... & ... & ... & ... & ... & ... & ... & ... & 17 \\
PS15aic & 2MASX J13304792+3806450 & Sc & 0.056 & ... & ... & ... & ... & 0.00 & 0.51 & 0.34 & ... & 17,18 \\
SN2015H$^d$ & NGC 3464 & Sc & 0.012 & -17.27(07) & 0.69(04) & ... & 0.42 & 0.71 & 0.89 & 0.87 & 8.50(10) & 14,16 \\
OGLE16erd & ... & ... & 0.068 & ... & ... & ... & ... & ... & ... & ... & ... & 17 \\
SN2016atw & ... & Sbc/Sc & 0.065 & ... & ... & ... & ... & ... & ... & ... & ... & 17 \\
SN2016ilf & 2MASX J02351956+3511426 & Sb & 0.045 & ... & ... & ... & ... & ... & ... & ... & ... & 15,18 \\
SN2017gbb$^d$ & IC 438 & Scd & 0.010 & $\leqslant$-17.00 & ... & ... & 0.52 & ... & ... & ... & ... & 17 \\
		\hline
        \multicolumn{13}{l}{$^a$ The host galaxy radius is taken from NASA/IPAC Extragalactic Database (NED).}\\
        \multicolumn{13}{l}{$^b$ Fractional fluxes in $u$ band, measured with SDSS DR12 and PanSTARRS images.}\\
        \multicolumn{13}{l}{$^c$ (1)This paper; (2)Foley et al. 2009; (3)Foley et al. 2013; (4)Li et al. 2003; (5)Phillips et al. 2007;
        (6)McClelland et al. 2010; (7)Foley et al. 2010; (8)Stritzinger et al. 2014;}\\
        \multicolumn{13}{l}{     (9)Szalai et al. 2015; (10)Yamanaka et al. 2015; (11)Magee et al. 2017;
        (12)Liu et al. 2015; (13)Tomasella et al. 2016; (14)Magee et al. 2016; (15)Transient Name Server;}\\
        \multicolumn{13}{l}{      (16)Lyman et al. 2017; (17)Guillochon et al. 2017;
        (18)Morphology judged from images on PanSTARRS; (19)White et al. 2015; (20)Jha 2017; (21)Jha et al. 2013.}\\
        \multicolumn{13}{l}{$^d$ PTF10ujn is classified as a 02es-like supernova in White et al. 2015. Magnitude for SN 2015H and SN 2017gbb is from SDSS r-band and g-band respectively.} \\
	\end{tabular}
    \end{sideways}
\end{table*}

\begin{table}
	\centering
	\caption{KS Test Results on the distribution of Fractional Flux}
	\label{tab:table 6}
	\begin{tabular}{cccc}
		\hline
		SN Type & $f_u$ & $f_g$ & $f_r$ \\
		\hline
		Ia & 0.03 & 0.19 & 0.04 \\
  Ibc & 0.05 & 0.33 & 0.39  \\
II & 0.03 & 0.13 & 0.17 \\
		\hline
	\end{tabular}
\end{table}

\bsp
\label{lastpage}
\end{document}